\documentclass{emulateapj}

\newcommand{\etal}{{et~al.}}

\shorttitle{EDisCS Morphological Content}
\shortauthors{Desai \etal}

\begin{document}

\submitted{Accepted for Publication in ApJ}

\title{The Morphological Content of Ten EDisCS Clusters at {\boldmath
$0.5 < z < 0.8$}\altaffilmark{1,2}}

\author{V.~Desai\altaffilmark{3,4}, 
J.~J.~Dalcanton\altaffilmark{4},
A.~Arag{\' o}n-Salamanca\altaffilmark{5},
P.~Jablonka\altaffilmark{6},
B.~Poggianti\altaffilmark{7}, 
S.~M.~Gogarten\altaffilmark{4},
L.~Simard\altaffilmark{8}, 
B.~Milvang-Jensen\altaffilmark{9},
G.~Rudnick\altaffilmark{10}, 
D.~Zaritsky\altaffilmark{11},
D.~Clowe\altaffilmark{11},
C.~Halliday\altaffilmark{12},
R.~Pell{\' o}\altaffilmark{13},
R.~Saglia\altaffilmark{14},
S.~White\altaffilmark{15}}

\altaffiltext{1}{Based on observations made with the NASA/ESA
\textit{Hubble Space Telescope}, obtained at the Space Telescope
Science Institute, which is operated by the Association of
Universities for Research in Astronomy, Inc., under NASA contract NAS
5-26555. These observations are associated with proposal 9476.
Support for this proposal was provided by NASA through a grant from
the Space Telescope Science Institute.}

\altaffiltext{2}{Based on observations obtained at the ESO Very Large
Telescope (VLT) and New Technology Telescope (NTT) as part of the
large program 166.A-0162 (the ESO Distant Cluster Survey).}

\altaffiltext{3}{Division of Physics, Mathematics and Astronomy,
320-47, California Institute of Technology, Pasadena CA 91125, USA}

\altaffiltext{4}{University of Washington Department of Astronomy, Box
351580, Seattle WA 98195-1580, USA}

\altaffiltext{5}{School of Physics and Astronomy, University of
Nottingham, Nottingham NG7 2RD, UK}

\altaffiltext{6}{Universit\'e de Gen\`eve, Laboratoire d'Astrophysique
 de l'Ecole Polytechnique F\'ed\'erale de Lausanne (EPFL),
 Observatoire, CH-1290 Sauverny, Switzerland}

\altaffiltext{7}{Osservatorio Astronomico di Padova, Vicolo
dell'Osservatorio 5, 35122 Padua, Italy}

\altaffiltext{8}{Herzberg Institute of Astrophysics, National Research
Council of Canada, 5071 West Saanich Road, Victoria, BC V9E 2E7,
Canada}

\altaffiltext{9}{Dark Cosmology Centre, Niels Bohr Institute,
University of Copenhagen, Juliane Maries Vej 30, DK-2100 Copenhagen,
Denmark }

\altaffiltext{10}{NOAO, 950 North Cherry Avenue, Tucson AZ 85719, USA}

\altaffiltext{11}{Steward Observatory, University of Arizona, 933 North
Cherry Avenue, Tucson, AZ 85721, USA}

\altaffiltext{12}{Osservatorio Astrofisico di Arcetri, Largo E. Fermi 5, 50125 Firenze, Italy}

\altaffiltext{13}{Laboratoire d'Astrophysique de l'Observatoire
Midi-Pyr\'en\'es, 14, Avenue Edouard Belin, F-31400 Toulouse, France}

\altaffiltext{14}{Max-Planck-Institut f\"{u}r extraterrestrische Physik,
Giessenbachstrasse, D-85748 Garching bei M\"{u}nchen, Germany}

\altaffiltext{15}{Max-Planck-Institut f\"{u}r Astrophysik,
Karl-Schwarzschild-Strasse 1, Postfach 1317, D-85748 Garching bei
M\"{u}nchen, Germany}

\begin{abstract}
We describe Hubble Space Telescope (HST) imaging of 10 of the 20 ESO
Distant Cluster Survey (EDisCS) fields.  Each $\sim$40 square
arcminute field was imaged in the F814W filter with the Advanced
Camera for Surveys Wide Field Camera.  Based on these data, we present
visual morphological classifications for the $\sim$920 sources per
field that are brighter than $I_{\rm auto}=23$ mag.  We use these
classifications to quantify the morphological content of 10
intermediate-redshift ($0.5 < z < 0.8$) galaxy clusters within the HST
survey region.  The EDisCS results, combined with previously published
data from seven higher redshift clusters, show no statistically
significant evidence for evolution in the mean fractions of
elliptical, S0, and late-type (Sp+Irr) galaxies in clusters over the
redshift range $0.5 < z < 1.2$.  In contrast, existing studies of
lower redshift clusters have revealed a factor of $\sim$2 increase in
the typical S0 fraction between $z=0.4$ and $z=0$, accompanied by a
commensurate decrease in the Sp+Irr fraction and no evolution in the
elliptical fraction.  The EDisCS clusters demonstrate that cluster
morphological fractions plateau beyond $z \approx 0.4$.  They also
exhibit a mild correlation between morphological content and cluster
velocity dispersion, highlighting the importance of careful sample
selection in evaluating evolution.  We discuss these findings in the
context of a recently proposed scenario in which the fractions of
passive (E,S0) and star-forming (Sp,Irr) galaxies are determined
primarily by the growth history of clusters.
\end{abstract}


\section{Introduction}
\label{sec:intro}

Although the morphology-density relation is observationally
well-established, it is still unclear why the incidence of early type
(E+S0) galaxies is higher in over-dense regions of the universe. The
options are often couched in terms of ``nature'' versus ``nurture''.
These scenarios describe different paths to qualitatively similar
relations between morphology and environment, and are therefore
difficult to disentangle.  In the nature scenario, galaxies that end
up in high density environments at low redshift are more likely to
have experienced initial conditions leading to an early-type
morphology {\em upon formation}.  In the ``nurture'' scenario,
all galaxies, regardless of their local densities at low redshift, have
identical probabilities of having formed as early types.  The
morphology-density relation is then the result of subsequent
morphological alterations as galaxies enter increasingly higher
density environments as structure grows.  Many mechanisms that could
produce the required morphological transformations have been
suggested, including mergers and galaxy-galaxy interactions
\citep{Toomre72,Icke85,Lavery88,Mihos04}, harassment
\citep{Richstone76,Moore98}, gas stripping \citep{Gunn72,Abadi99,
Quilis00}, strangulation \citep{Larson80,Bekki02}, and cluster tidal
forces \citep{Byrd90}.

The observed evolution of the morphology-density relation provides an
important constraint on models of its origin.  Observations of the
high-density regions of cluster cores show that while the elliptical
fraction has not evolved, the S0 (Sp+Irr) fraction has grown
(diminished) by a factor of $\sim$2 over the last 5 Gyr
\citep{Dressler97, Fasano00}.  However, over the same time interval,
no evolution is observed at lower densities \citep{Treu03,Smith05}.
The morphology-density relation at larger lookback times ($z > 0.5$)
has just begun to be explored.  \citet{Smith05} used Hubble Space
Telescope (HST) Wide-Field Planetary Camera 2 (WFPC2) data for six
clusters at $0.75 < z < 1.25$ to estimate the early-type fraction as a
function of local galaxy surface density.  They extended their
measurements to low densities ($\sim$1 Mpc$^{-1}$) by including field
galaxies in HST WFPC2 images of the cluster CL0024 at $z = 0.395$.
Comparing their results to similar studies at lower redshifts
\citep{Dressler80b, Dressler97, Treu03}, they found that the
early-type fraction has increased steadily since $z \sim 1$ in the
densest regions ($\sim$1000 Mpc$^{-2}$), has increased only since $z
\sim 0.5$ in moderate density regions ($\sim$100 Mpc$^{-2}$), and has
remained constant since $z \sim 1$ in the lowest density regions
($<$10 \rm Mpc$^{-2}$).  \citet{Postman05} find compatible results in
their study of the morphology-density relation using Advanced Camera
for Surveys (ACS) HST imaging of seven $z \sim 1$ clusters.
\citet{Smith05} suggest a nurture scenario to explain the observed
evolution in the morphology-density relation.  If structure growth is
hierarchical, the densest regions at any redshift collapsed the
earliest.  Thus, if environmental processes can modify a universal
initial mix of morphologies through a transformation from late to
early types, these processes would have had a longer time to operate
in increasingly dense regions.  In addition, the efficiency of such
transformations could be density-dependent.

To discover whether environmental processes, rather than initial
conditions, drive the differential evolution of the morphology-density
relation, we must first demonstrate that nature scenarios cannot
reproduce the observations.  If we find that environmental factors are
indeed important, it is critical to identify the responsible
mechanisms.  The current data are insufficient to discriminate among
the alternative models.  For this, we require a full mapping between
galaxy properties and environment.  Towards this goal, we present the
morphological fractions in ten rich clusters at $0.5 < z < 0.8$, drawn
from the ESO Distant Cluster Survey\footnote{{\tt
http://www.mpa-garching.mpg.de/galform/ediscs/index.shtml}}
\citep[EDisCS;][]{White05}.  Because strong evolution has been observed
in high-density regions, clusters are a critical environment to probe.
The redshift range of our study bridges a gap between those of
\citet{Dressler80a}, \citet{Fasano00}, and \citet{Dressler97} at low
redshift and those of \citet{Smith05} and \citet{Postman05} at high
redshift.  The EDisCS clusters represent a significant increase in the
number of well-studied clusters at high redshift.  Because the mean
age of clusters decreases with increasing redshift, the scatter in
cluster properties may also increase with redshift.  Thus, large
samples are necessary to disentangle evolution from cluster-to-cluster
variations.  Finally, we measure morphologies from high-quality HST ACS
images, allowing us to distinguish S0s from other early types.  As
described above, the S0 population appears to have increased at the
expense of the Sp+Irr population since $z = 0.5$.  Thus, tracing the
S0 population to high redshifts may place important constraints on
their formation.

This paper is organized as follows.  In \S{\ref{sec:EDisCSSample}} we
describe the EDisCS sample.  In \S{\ref{sec:HSTACSData}} we describe
the HST ACS data used to measure morphologies.  Our classification
procedure and Hubble types for all galaxies with $I_{\rm auto} < 23$
mag are presented in \S{\ref{sec:GalaxyMorphologies}}.  In
\S{\ref{sec:Analysis}} we describe our method of quantifying the
morphological content in each cluster.  Our results are presented in
\S{\ref{sec:Results}}, and our conclusions regarding them can be found
in \S{\ref{sec:Conclusions}}.

Results will be presented for two sets of cosmological parameters, to
allow a direct comparison with previous work ($\Omega_0 = 1$, $\Lambda
= 0$, H$_0 = 50$ km s$^{-1}$ Mpc$^{-1}$) and to provide the
corresponding estimate for the currently standard cosmology ($\Omega_0
= 0.3$, $\Lambda = 0.7$, H$_0 = 70$ km s$^{-1}$ Mpc$^{-1}$).  In the
remainder of this paper we will refer to these as the classic
cosmology and the WMAP cosmology, respectively.

\section{The ESO Distant Cluster Survey (EDisCS)}
\label{sec:EDisCSSample}

The ten clusters included in this study are a subset of the ESO
Distant Cluster Survey, an extensive program to obtain
spectroscopy and multiwavelength photometry for 20 galaxy clusters at
$0.4 < z < 1$.  Details of the goals and data sets associated with
this survey are documented by \citet{White05}, but we briefly
summarize below.

The clusters chosen to be part of EDisCS were drawn from the Las
Campanas Distant Cluster Survey (LCDCS) catalog, which consists of
optically-selected clusters identified from fluctuations in the
extragalactic background light \citep{Dalcanton96, Zaritsky97,
Gonzalez01}.  The redshifts of the LCDCS cluster candidates were
initially estimated from the apparent magnitude of the brightest
cluster galaxy (BCG).  The candidates were divided into an
intermediate-redshift group ($z \sim 0.5$) and a high-redshift group
($z \sim 0.8$).  Shallow two-color imaging was obtained at the Very
Large Telescope (VLT) for the 30 brightest candidates in each group.
A sample of 20 EDisCS clusters was selected based on the presence of a
red sequence and a BCG of the appropriate magnitude
\citep{Gonzalez01}.  These twenty clusters were then imaged more
deeply at the VLT in BVI (45 minutes) for the $z \sim 0.5$ candidates
and in VRI (2 hours) for the $z \sim 0.8$ candidates.  These data,
covering a 6.5$\arcmin$ $\times$ 6.5$\arcmin$ region in the field of
each cluster, are presented in \citet{White05}, and form the basis of
the weak-shear analysis presented by \citet{Clowe06}.  Additional
near-infrared imaging was obtained with SOFI at the New Technology
Telescope (NTT) in $K_s$ for the $z \sim 0.5$ candidates and in $JK_s$
for the $z \sim 0.8$ candidates (Arag\'{o}n-Salamanca \etal, in
preparation).  The near-infrared imaging covers approximately
6$\arcmin$ $\times$ 4.2$\arcmin$ in the fields of the
intermediate-redshift clusters and 5.4$\arcmin$ $\times$ 4.2$\arcmin$
in the fields of the high-redshift clusters.  The optical and
near-infrared imaging were used to measure photometric redshifts, as
described in Pell{\' o} \etal \ (in preparation).

An initial phase of spectroscopy consisted of relatively short
exposures of a single slit mask per cluster.  One cluster with an
estimated redshift of $z \sim 0.8$ was revealed as a superposition of
weak groups, and was rejected from the sample.  These spectra revealed
that the remaining 19 clusters have redshifts in the range $0.4 \la z
\la 1$.  Deeper spectroscopic exposures ($\sim$2 hours each for
galaxies with $18.6 \le I(r=1\arcsec) \le 22$ mag in
intermediate-redshift clusters; $\sim$4 hours each for galaxies with
$19.5 \le I(r=1\arcsec) \le 23$ mag in high-redshift clusters) of 3-5
masks per field were then taken, and are of sufficient quality to
obtain information about the stellar populations and internal dynamics
of the target galaxies.  The spectroscopic observations are described
in \citet{Halliday04} and Milvang-Jensen \etal \ (in preparation).

The ground-based EDisCS imaging and spectroscopy go a long way towards
characterizing both the clusters themselves and the galaxies within
them.  At the high redshifts of the EDisCS sample, however, only HST
can provide the spatial resolution necessary to provide robust
morphologies.  Motivated by the issues discussed in
\S{\ref{sec:intro}}, we obtained HST imaging for the ten highest
redshift EDisCS clusters, listed in Table \ref{table:HSTSample} along
with their basic physical parameters.  Cluster 9 (cl1232-1250) was
drawn from the EDisCS intermediate-redshift sample, and the remainder
were drawn from the high-redshift sample.

\section{HST ACS Data}
\label{sec:HSTACSData}

The HST observations were designed to coincide as closely as possible
with the coverage of the ground-based optical imaging and
spectroscopy, within guide star constraints.  The ground-based optical
data cover a 6.5$\arcmin$ $\times$ 6.5$\arcmin$ region around each
cluster, with the cluster center displaced by 1$\arcmin$ from the
center of the region.  For reference, the ACS Wide Field Camera has a
field of view of roughly 3.5$\arcmin$ $\times$ 3.5$\arcmin$.
Balancing scientific motives for going deep over the entire field
against a limited number of available orbits, we tiled each
6.5$\arcmin$ $\times$ 6.5$\arcmin$ field in four pointings, with one
additional deep pointing on the cluster center (taken as the location
of the BCG).  The resulting exposure time per pixel is 2,040 seconds,
except for the central 3.5$\arcmin$ $\times$ 3.5$\arcmin$, which has
an exposure time per pixel of 10,200 seconds.  The deep central
pointing probes to lower surface brightness, fainter magnitudes, and
larger galactic radii in the region of the cluster containing the most
galaxies.  The centers of the structures ultimately chosen for
spectroscopic followup in the cl1103-1245b and cl1227-1138 fields are
somewhat offset from those anticipated at the time the HST
observations were taken.  All exposures were taken under LOW SKY
conditions to maximize our surface brightness sensitivity.

The ACS calibration pipeline, {\sc CALACS v4.3} (6-June-2003)
debiased, dark-subtracted, and flat-fielded our ACS images ``on the
fly'' when they were requested from the HST archives.  Known bad
pixels and saturated data were also flagged in accompanying data
quality (DQ) arrays.  Approximate World Coordinate System (WCS)
headers were provided.  The resulting images were returned by the
archive with the FLT suffix.  In addition, images that were cosmic-ray
split were combined and returned with a CRJ suffix.

To produce a mosaic, we required precise offsets between the 32 FLT
images retrieved for each cluster.  The approximate WCS headers
provided were insufficient for this purpose.  We found that the
offsets between cosmic-ray split images were negligible.  We therefore
computed the necessary shifts between pointings using the higher
signal-to-noise CRJ images.  The shifts could not be computed using
the retrieved CRJ files because they had not been undistorted.  We
therefore drizzled each CRJ image separately, using the approximate
WCS headers provided.  Because each undistorted CRJ image overlaps the
central pointing, one of the images centered on the cluster center was
chosen as the reference image.  Shifts between the reference image and
each undistorted CRJ image were then computed using cross-correlation.

Combination of the FLT images using the resulting shifts was
accomplished using {\it MultiDrizzle}, a Python code written by Anton
Koekemoer to run under PyRAF, the Python-based interface to the Image
Reduction and Analysis Facility (IRAF).  {\it MultiDrizzle}
automatically removes cosmic rays and combines dithered images using
{\it PyDrizzle}, which has been developed by the Science Software
Branch at the Space Telescope Science Institute.  For each of the 32
FLT images per cluster, {\it MultiDrizzle} included negative bad
pixels in the data quality array, subtracted the sky, and separately
drizzled and undistorted each image.  Next, it created a median image
from these separately drizzled images using shifts computed from the
headers along with the user-supplied refinement shifts described
above.  This median image, relatively free of cosmic rays, was
compared to the input images to identify cosmic rays.  The median
image was then re-distorted to create cosmic ray masks.
These masks were used in the final image combination step using {\it
drizzle} and the {\it lanczos3} kernel, which provided optimal noise
properties.

\section{Galaxy Morphologies}
\label{sec:GalaxyMorphologies}

We visually classified all galaxies brighter than $I_{\rm auto} = 23$
mag.  Here $I_{\rm auto}$ is the SExtractor \citep{Bertin96} AUTO
magnitude measured on the I-band VLT images, and is an estimate of the
total magnitude of a galaxy, in the Vega system.  This magnitude does
not include an aperture correction, nor has it been corrected for
Galactic extinction.  The limit was set both to ensure robust
classifications and to provide a tractable sample of $\sim$9200
galaxies.

Our classifications are most useful if they conform to systems adopted
by previous studies.  For this reason, each classifier trained on the
HST WFPC2 images and visual morphological catalogs of the $0.3 < z <
0.5$ MORPHS clusters, using the same procedure described in
\cite{Smail97}.  For uniformity, each classifier used the same IRAF
script to examine and classify EDisCS galaxies.  This script displays
two side-by-side versions of a 200 $\times$ 200 square pixel cutout centered on
each galaxy meeting the magnitude limit described above.  One version
is on a log scale between -0.1 -- 2 DN/s, while the other is on a log
scale between -0.1 -- 25 DN/s.  Together, these displays allowed
classifiers to inspect the galaxies from their high surface brightness
cores to their low surface brightness outer features.  In general, we
found that the depth and quality of the EDisCS ACS data were similar
to or better than the MORPHS WFPC2 data for lower redshift clusters.

Classifications were performed by 5 of the authors (AAS, JJD, VD, PJ,
BP).  Each classified the galaxies in 3--6 clusters.  For a given
galaxy, the final Hubble type was based on the classifications of two
or more of the authors.  First, the Revised Hubble Type was translated
into a T-type according to the scheme presented in Table
\ref{table:parameters}.  Classifications appended by one question mark
were given half weight; those appended by two question marks were
given one-quarter weight.  If a classifier specified two types
separated by a slash, the first was given three-quarters weight and
the second was given one-quarter weight.  All T types submitted for a
given galaxy were then ranked by their frequency.  The final Hubble
Type was then chosen from the highest ranked T types. In general, the
highest ranked T type was chosen.  If there were two equally-ranked
classifications, one was chosen randomly.  If there were three or more
equally-ranked classifications, the differences in T types were
computed between them.  If there were no gaps smaller than three T
types, the median value was chosen.  If there was one gap smaller than
three T types, one of the two similar classifications was chosen at
random.  If there was more than one gap smaller than three T types,
the median value of all classifications was chosen.

The main results of this work are sensitive to how accurately galaxies
can be placed into the broad categories of elliptical (E), S0, and
late (Sp+Irr).  Although accuracy in this case is difficult to
quantify, we can test how consistently galaxies are placed in the same
bin by different classifiers.  Using $\sim$900 galaxies down to
$I_{\rm auto}=23$ mag in cl1216-1201, all of which were assigned a
Hubble type by all classifiers, we computed the raw fractions for each
classifer, uncorrected for the presence of foreground and background
galaxies.  The root mean variance for any morphological fraction is
$\la 0.10$, which is comparable to the error computed from Poisson
statistics.

\begin{figure*}[h]
\epsscale{0.8}
\plotone{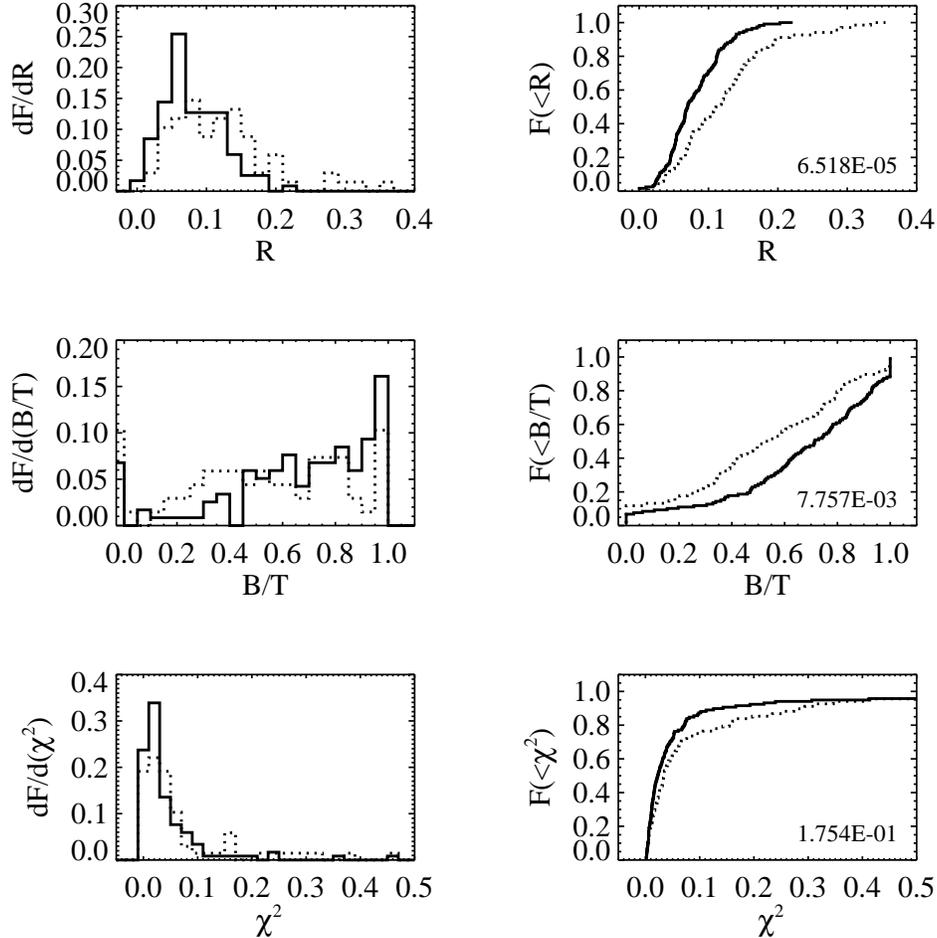}
\caption{Comparison of parameters derived from 1D profile fits to
visually-classified ellipticals and S0s down to $I_{\rm auto} = 23$
mag in cl1216-1201.  From top to bottom, this figure shows the
fractional differential (left) and cumulative (right) distribution
functions of the width of the ellipticity distribution within a given
galaxy (R), the bulge-to-total fraction (B/T), and the $\chi^2$
goodness-of-fit parameter for a pure de Vaucouleurs component (see \S
\ref{sec:GalaxyMorphologies}).  The solid lines indicate the
distributions for ellipticals, and the dotted lines indicate the
distributions for S0 galaxies.  The significance levels of the
two-sided Kolmogorov-Smirnov tests comparing the E and S0
distributions are shown in the bottom right of the cumulative
distribution plots.  The R and B/T distributions for Es and S0s are
different at very high significance levels, while the $\chi^2$
goodness-of-fit parameters are different at a low significance level.
These results show that galaxies that are classified as Es are
structurally distinct from those classified as S0s.}
\label{fig:profiles} 
\end{figure*}

There is some controversy surrounding the ability of morphologists to
distinguish between E and S0 galaxies, especially at high redshifts.
For this reason, some previous investigators chose to lump E and S0
galaxies together into an early-type class.  For easy comparison with
these works, in the following we have plotted and tabulated our
results for early types as well as for E and S0 galaxies separately.
However, because the S0 population is strongly evolving in clusters at
$z < 0.5$, it is important to track this population at higher
redshifts.  Comparing the EDisCS ACS data to the MORPHS WFPC2 data
gives the impression that we can distinguish S0s from ellipticals with
the same accuracy for the EDisCS sample as was obtained with the
MORPHS sample.  To test this impression, we have performed a
quantitative analysis of the light distributions of E and S0 galaxies
within cl1216-1201, which is our highest-redshift cluster and which
has been visually examined by all classifiers.  Specifically, we used
the ELLIPSE task in the Space Telescope Science Data Analysis System
(STSDAS) package of IRAF to measure the surface brightness of a series
of isophotal ellipses in each galaxy as a function of semi-major axis.
For each step, the semi-major axis is increased by a factor of 1.1.
The ELLIPSE task was allowed to vary the ellipticity and position
angle of the ellipses within a given galaxy.  For each surface
brightness profile, we fit a bulge+disk model of the form

\begin{equation}
I_{\rm bulge}(r) = I_{\rm e} \exp{(-7.67[(r/r_{\rm e})^{1/4} - 1])}
\end{equation}

\noindent and

\begin{equation}
I_{\rm disk}(r) = I_{\rm 0} \exp{(-r/r_{\rm d})},
\end{equation}

\noindent where the total surface brightness is given by $I_{\rm
bulge} + I_{\rm disk}$.  We considered three parameters for which E
and S0 galaxies should display different distributions: 1) the width
of the ellipticity distribution in a given galaxy, $R$, defined as the
interquartile range of all values of the ellipticity for a given
galaxy; 2) the bulge-to-total fraction B/T; and 3) $\chi^2$
goodness-of-fit from fitting the $r^{1/4}$ law (i.e. only bulge, no
disk).  The differential and fractional cumulative distribution
functions for these three parameters are plotted in Figure
\ref{fig:profiles}.  At a statistically significant level, S0s display
larger ellipticity distribution widths than ellipticals.  The presence
of a disk in a galaxy will result in a wider range of ellipticities,
since the ellipticity of the galaxy changes as the disk begins to
dominate the surface brightness profile.  Likewise, the B/T fractions
for visually-classified ellipticals appear to be skewed to higher
values compared to S0s, as expected.  This is confirmed by a two-sided
Kolmogorov-Smirnov test.  Finally, a higher fraction of ellipticals
have small values of $\chi^2$, indicating a better fit to a pure bulge
model.  However, the differences in the $\chi^2$ distribution are not
significant.  The overall results of these tests are that
visually-classified S0s have quantitatively different surface
brightness profiles than visually-classified ellipticals.  Although in
some individual cases it is difficult for visual classifiers to
differentiate between Es and S0s, it is clear that, in a statistical
sense, objects visually classified as Es and S0s form two distinct
populations with objectively measurable physical differences.  More
sophisticated two-dimensional profile fits will be presented in Simard
et al., in preparation.

Morphological catalogs for the ten highest-redshift EDisCS clusters
are available in the on-line version of this paper, with columns
described in Table \ref{table:parameters}.

\section{Analysis}
\label{sec:Analysis}

The goal of this paper is to quantify the morphological content of the
EDisCS clusters.  In particular, we will discuss the overall fractions
of E, S0, and late-type galaxies in our clusters and compare them to
the fractions found in clusters spanning a range of redshifts.
Following \citet{Dressler97}, we do not include galaxies that were
unclassifiable (types -6 and 66 in Table \ref{table:parameters}) in
our analysis.  To facilitate a fair comparison with other samples, we
compute the morphological fractions as consistently as possible with
previous work.  In the following subsections, we describe the key
elements of our analysis.

\subsection{Magnitude Range}
\label{sec:MagnitudeRange} 

Early types preferentially occupy the bright end of the galaxy
luminosity function, while late types dominate the faint end
\citep[e.g.][]{Blanton01,Goto02,Zucca06}.  Thus, morphological
fractions depend upon the range of absolute magnitudes sampled.
Morphological fractions in the MORPHS clusters were determined using
apparent magnitude cuts in $I_{\rm 814}$ designed to correspond to a
$V$-band absolute magnitude of $M_V = -20$ mag \citep{Dressler97}, but
in actuality corresponding to $M_V = -19$ mag due to a transcription
error \citep{Fasano00}.  However, \citet{Fasano00} re-analyzed the
MORPHS data with the intended limiting absolute magnitude and
additionally analyzed nine clusters with $0.1 < z < 0.25$ in the same
manner, providing an ideal comparison sample for the EDisCS clusters.
We therefore adopt an absolute magnitude limit of $M_V = -20$ mag.
For a given cluster, the limiting apparent magnitude in the $I$-band
is then given by

\begin{equation}
I_{{\rm auto, lim}} = M_{\rm V} + 5 \log_{10}(d_{\rm L,pc}) - 5 -
(M_{\rm V} - M_{\rm I}) + k_{\rm I},
\label{equation:faintlim}
\end{equation}

\noindent where $d_{\rm L,pc}$ is the luminosity distance of the
cluster in parsecs (calculated in either the classic or WMAP
cosmology, as indicated), $M_{\rm V} - M_{\rm I}$ is the rest-frame
color, and $k_{\rm I}$ is the $I$-band k-correction.  We adopt the
rest-frame color and Cousins $I$-band k-corrections for an elliptical
galaxy, as presented in \citet{Poggianti97}.  Equation
\ref{equation:faintlim} results in values of $I_{{\rm auto, lim}}$
ranging from 21.3 to 23.3 mag, the faintest values being slightly
fainter than the limit of our visual classifications.  Given the
errors in the k-corrections relative to these differences, we adopt
$I_{{\rm auto, lim}} = 23$ mag in these instances.  This occurs for
cl1216-1201 when using the WMAP cosmology and for cl1054-1245,
cl1216-1201, and cl1354-1230 when using the classic cosmology.

\subsection{Aperture}
\label{sec:Aperture} 

Clusters exhibit a radial gradient in their morphological fractions;
the centers of clusters contain a larger fraction of ellipticals than
the outskirts \citep[e.g.][]{Melnick77,Goto04,Thomas06}.  Thus, in
comparing the overall morphological fractions across clusters, one
must choose a consistent aperture.  The MORPHS project has set the
standard for the aperture within which to compute morphological
fractions. Most studies of the morphological fractions of galaxy
clusters at other redshifts have used a similarly-sized aperture. So
that we may assess the level of evolution between the morphological
studies presented in the literature and those conducted with EDisCS
data, we adhere to this precedent and adopt a circular aperture of
radius 600 kpc (classic cosmology).

Since the radial density profiles of clusters vary, the average galaxy
density within a fixed metric aperture will also vary.  We therefore
also calculated the morphological fractions within an aperture that
scales with $R_{200}$, the radius within which the average mass
density is equal to 200 times the critical density.  The derived
$R_{200}$ values were computed using Equation 8 in
\citet{Finn05}, and are listed in Table \ref{table:HSTSample} for
both the classic and WMAP cosmologies.  Unfortunately, our imaging
data is complete out to $R_{200}$ for only cl1040-1155, cl1054-1146,
cl1054-1245, and cl1354-1230.  We therefore use a radius of 0.6
$\times$ $R_{200}$, which keeps the analysis area within the imaging
region for all clusters except for cl1227-1138 and cl1232-1250.  The
fraction of the analysis region which is not included in these two
clusters is very small, and the resulting effect on the morphological
fractions is likely minimal.

\subsection{Background Subtraction}
\label{sec:BackgroundSubtraction}

Galaxies within the magnitude range and distance from the cluster
center described in \S{\ref{sec:MagnitudeRange}} and
\S{\ref{sec:Aperture}} lie at a variety of redshifts.  We wish to
determine the proportions of E, S0, and late-type galaxies of cluster
members only.  Because the morphological mix of field galaxies differs
substantially from that of cluster populations, morphological
fractions computed without regard to field contamination will
underestimate the fraction of early types and overestimate the
fraction of late types in clusters.  We use four methods, described
in detail below, to account for the presence of cluster non-members,
checking for consistency among the resulting morphological fractions.

\subsubsection{Spectroscopic Redshifts}
\label{sec:SpectroscopicRedshifts}

As described in \S{\ref{sec:EDisCSSample}}, the EDisCS program
includes an extensive spectroscopic survey.  For each cluster,
redshifts were obtained for 6-86 galaxies which both meet our
apparent magnitude limit and lie within $0.6 \times R_{200}$
(WMAP cosmology).  We use this spectroscopic sample to constrain the
morphological fractions in each cluster in two ways.  First, we
calculate hard limits on the morphological fractions using both the
spectroscopic sample and the purely photometric sample, consisting of
galaxies that were either not targeted for spectroscopy, or for which
spectroscopy failed to yield a redshift.  This calculation does not
require that the spectroscopic sample be complete.  Second, we
estimate the morphological fractions using only the spectroscopic
sample after applying small completeness corrections.

\begin{figure*}[h]
\epsscale{1.0}
\plotone{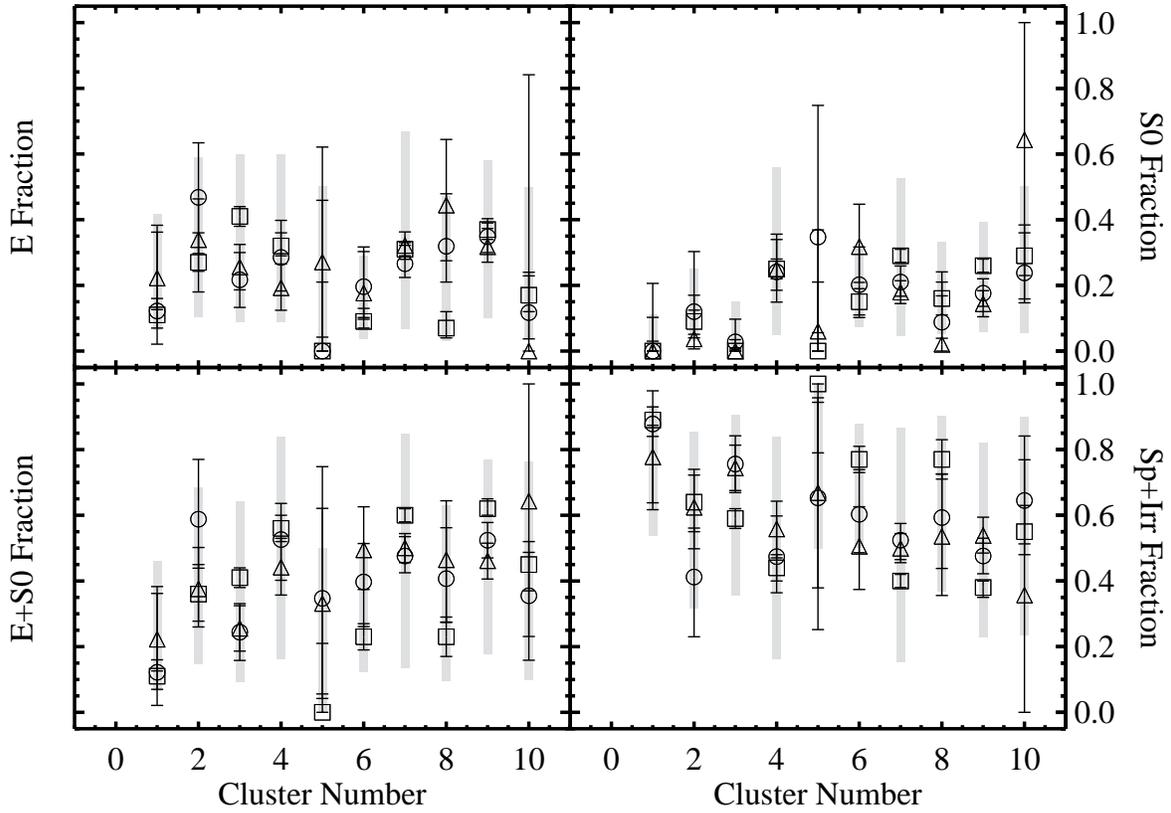}
\caption{A comparison of four methods for estimating the E, S0, Sp,
and Sp+Irr fractions in the 10 EDisCS clusters listed in Table
\ref{table:HSTSample}.  The shaded grey regions span the full range
allowed by our spectroscopic data, while the squares show the best
estimates of the morphological fractions determined using
spectroscopic redshifts (see \S \ref{sec:SpectroscopicRedshifts}).
Circles and triangles represent the morphological fractions determined
using photometric redshifts (see \S \ref{sec:PhotometricRedshifts})
and statistical background subtraction (see \S \ref{sec:StatSub}),
respectively.  All calculations were performed using the WMAP
cosmology and an aperture of 0.6 * $R_{200}$.}
\label{fig:MethodComparison} 
\end{figure*}

The upper and lower hard limits on the fraction of galaxies of type
$i$ are given by:

\begin{equation}
f_{\rm min}(i) = \frac{N_{\rm s}^{\rm m}(i)}{N_{\rm s}^{\rm m}
+ N_{\rm p}^{\rm u} - N_{\rm p}^{\rm u}(i)}
\label{equation:spec1}
\end{equation}

\noindent and 

\begin{equation}
f_{\rm max}(i) = \frac{N_{\rm s}^{\rm m}(i) + N_{\rm
p}^{\rm u}(i)}{N_{\rm s}^{\rm m} + N_{\rm p}^{\rm u}(i)},
\label{equation:spec2}
\end{equation}

\noindent where the subscripts $s$ and $p$ refer to the spectroscopic
sample and the photometric sample, respectively; and the superscripts
$m$ and $u$ indicate cluster members and galaxies at unknown
(spectroscopic) redshift, respectively.  Each quantity is a function
of galaxy type $i$.  The fraction of galaxies of type $i$ is minimized
when all of the other types that do not have redshifts are members and
all of the galaxies of type $i$ that do not have redshifts are
nonmembers.  The fraction of galaxies of type $i$ is maximized when
all of the galaxies of other types without redshifts are nonmembers
and all of the galaxies of type $i$ without redshifts are members.
The results of applying equations \ref{equation:spec1} and
\ref{equation:spec2} are shown as the grey shaded regions in Figure
\ref{fig:MethodComparison}.

To directly estimate the morphological fractions from the spectroscopic data,
we must correct for incompleteness in the spectroscopic sample.
As described in the Appendix of \cite{Poggianti06}, the completeness
of the EDisCS spectroscopy is a weak function of both apparent
$I$-band magnitude and clustercentric distance.  Using the
cluster-by-cluster magnitude and geometric completeness weights
$W_{\rm mag}$ and $W_{\rm geo}$ calculated in that work, we estimate
the spectroscopic morphological fractions $f_{\rm spec}(i)$ as:

\begin{equation}
N_{\rm spec}(i) = \sum_{j=1}^{N_{\rm s}^{\rm m}(i)} W_{\rm mag}(j)^{-1} W_{\rm geo}(j)^{-1} 
\end{equation}

\begin{equation}
N_{\rm spec}(tot) = \sum_{j=1}^{N_{\rm s}^{\rm m}} W_{\rm mag}(j)^{-1} W_{\rm geo}(j)^{-1}.
\end{equation}

\begin{equation}
f_{\rm spec}(i) = N_{\rm spec}(i) / N_{\rm spec}(tot)
\label{equation:spec3}
\end{equation}

The probability of measuring a given morphological fraction is the
product of the Poisson probability of measuring the total number of
cluster members times the binomial probability of measuring the
observed number of a given morphology.  We use the approximation
presented in Equations 21 and 26 of \citet{Gehrels86} to derive
1-$\sigma$ error estimates based on this premise, using $N_{\rm
spec}(i)$ and $N_{\rm spec}(tot)$ as inputs.  The results of applying
Equation \ref{equation:spec3} are shown as hollow squares in Figure
\ref{fig:MethodComparison}.

As described in \S \ref{sec:EDisCSSample}, the faint magnitude limit
of the EDisCS spectroscopic survey is $I(r = 1\arcsec) = 22$ mag for
the intermediate-redshift clusters and $I(r = 1\arcsec) = 23$ mag for
the high-redshift clusters.  In general, $I(r = 1\arcsec)$ is fainter
than $I_{\rm auto}$.  For six of our clusters, all of the galaxies
meeting the $I_{\rm auto}$ magnitude and aperture requirements of our
analysis have values of $I(r = 1\arcsec)$ brighter than the
spectroscopic survey limit.  However, for cl1054-1245, cl1216-1201,
cl1232-1250, and cl1354-1230, 10-20\% are fainter than the
spectroscopic flux limit.  Most ($\ga$80\%) of these are late-type
galaxies.  Thus, it is possible that our direct spectroscopic
estimates are biased slightly towards low late-type fractions.  In
Figure \ref{fig:MethodComparison}, we compare the morphological
fractions derived from different methods, and find that for
cl1054-1245, cl1216-1201, and cl1232-1250, the spectroscopic method
does yield lower late-type fractions than the photometric redshift or
statistical background subtraction methods.  However, all methods
produce late-type fractions that are consistent with one another,
within the errors.  For cl1354-1230, the spectroscopic method results
in a late-type fraction between the photometric redshift method and
the statistical background method.

\noindent 
\subsubsection{Photometric Redshifts}
\label{sec:PhotometricRedshifts}

Our optical and near-infrared imaging allows the derivation of
photometric redshifts, as described in Pell\'{o} \etal \ (in
preparation).  Briefly, two estimates of the redshift probability
distribution ($P(z)$) were computed for each galaxy.  Two independent
codes were employed, one described in \citet{Rudnick01} and
\citet{Rudnick03} and {\it Hyperz}, described in \citet{Bolzonella00}.
An estimate of $P(z)$ can be integrated over a suitable interval
$\Delta z$ (in this case $\pm$0.1) around the cluster redshift to
obtain the probability $P_{\rm clust}$ that the galaxy is a member of
the cluster.  We used our large spectroscopic sample to determine
$P_{\rm thresh}$, the minimum value of $P_{\rm clust}$ required for a
galaxy to be considered a cluster member.  Membership information
derived from each of our two estimates of $P(z)$ was then combined to
determine cluster membership.  That is, both estimates were required
to be consistent with cluster membership in order for a galaxy to be
considered a cluster member.

We used the subset (vast majority) of our spectroscopic sample with
both optical and near-infrared imaging to estimate a) the fraction of
spectroscopically-confirmed cluster members excluded by our
photometric redshifts as a function of morphology and b) the fraction
of spectroscopic non-members that are photometric members as a
function of morphology.  Since $P_{\rm thresh}$ was calibrated to
include as many cluster members as possible rather than to exclude all
non-members, the former fraction tends to be significantly smaller
than the latter.  Given these fractions, we calculated $N_{\rm m}(i)$,
the expected number of cluster members of type $i$ that were missed by
the photometric redshifts; and $N_{\rm c}(i)$, the expected number of
non-members among galaxies of type $i$ that contaminate the
photometric redshift member sample.  We then used this information to
compute the corrected number of observed cluster members of type $i$
($N_{\rm photoz}(i)$), and finally, the corrected fraction of cluster
members of type $i$ ($f_{\rm photoz}(i)$):

\begin{equation}
N_{\rm photoz}(i) = N_{\rm obs}(i) + N_{\rm m}(i) - N_{\rm
c}(i)
\end{equation}

\begin{equation}
N_{\rm photoz}(tot) = N_{\rm photoz}(E) + N_{\rm
photoz}(S0) + N_{\rm photoz}(Sp+Irr)
\end{equation}

\begin{equation}
f_{\rm photoz}(i) = N_{\rm photoz}(i) / N_{\rm photoz}(tot)
\label{equation:photoz}
\end{equation}

As in \ref{sec:SpectroscopicRedshifts}, errors were computed using the
\citet{Gehrels86} approximation using N$_{\rm photoz}(i)$ and N$_{\rm
photoz}(tot)$.  The morphological fractions computed using Equation
\ref{equation:photoz} are shown as hollow circles in Figure
\ref{fig:MethodComparison}.

As described in \S \ref{sec:EDisCSSample}, the area of each cluster
with near-infrared imaging is somewhat smaller than that imaged in the
optical.  In the classic cosmology, an analysis radius of 600 kpc
extends beyond the near-infrared imaging for cl1227-1138.  As a
result, 19 out of 49 galaxies meeting the magnitude and aperture
requirements for inclusion in the morphological analysis lack
near-infrared data.  Using the WMAP cosmology and an analysis radius
of 0.6 $\times$ $R_{200}$, a similar problem occurs for both
cl1227-1138 (21 out of 50 galaxies lack near-infrared data) and
cl1232-1250 (14 out of 222 galaxies lack near-infrared data).  Because
the $P(z)$ distributions for photometric redshifts computed without
near-infrared data are broad, they have low $P_{\rm clust}$ values,
and are somewhat more likely to be rejected than galaxies with
photometric redshifts computed using the full filter set.  Figure
\ref{fig:MethodComparison} shows that this effect does not appear to
have systematically skewed the morphological fractions for cl1227-1138
and cl1232-1250 compared to the other methods employed.

\subsubsection{Statistical Background Subtraction}
\label{sec:StatSub} 

In the absence of spectroscopic or photometric redshift information
for each galaxy, we can still estimate the morphological fractions by
statistically correcting the number of observed galaxies of a given
type to account for sources that lie in the field:

\begin{equation}
{\rm N}_{\rm field} = \Sigma_{\rm field} A,
\end{equation}

\begin{equation}
N_{\rm stat}(i) = N_{\rm obs}(i) - N_{\rm field} P(i)
\end{equation}

\begin{equation}
N_{\rm stat}(tot) = N_{\rm obs}(tot) - N_{\rm field}
\end{equation}

\begin{equation}
f_{\rm stat}(i) = N_{\rm stat}(i) / N_{\rm stat}(tot).
\label{equation:statsub}
\end{equation}

\noindent Here, $A$ is the area of the aperture described in
\S{\ref{sec:Aperture}}; N$_{\rm obs}(tot)$ is the total number of
galaxies meeting the magnitude criterion
(\S{\ref{sec:MagnitudeRange}}) within that aperture; N$_{\rm obs}(i)$
is the number of these galaxies which have morphology $i$; N$_{\rm
field}$ is the number of the observed galaxies that are expected to be
field members; and N$_{\rm stat}(tot)$ and N$_{\rm stat}(i)$ are the
background-subtracted number of galaxies in the aperture and the
number of type $i$, respectively.  The surface density of field
galaxies, $\Sigma_{\rm field}$, is determined by integrating the
$I$-band differential number counts in Table 1 of \citet{Postman98}
down to the redshift-dependent $I$-band apparent magnitude limit
adopted for each cluster, as described in \S \ref{sec:MagnitudeRange}.
We computed $P(i)$, the fraction of field galaxies of each
morphological type, using data from the Medium Deep Survey
\citep[MDS;][]{Griffiths94}, a Hubble Telescope Key Project that
cataloged the morphologies of intermediate-redshift field galaxies
down to $I_{\rm 814} \sim 22$ mag.  In particular, we used the
classifications of Richard Ellis listed in Table 1 of
\citet{Abraham96}.  The magnitude limits, $I_{{\rm auto, lim}}$, of
our morphological analysis vary from cluster to cluster and with
cosmology, but generally range from 21.3 to 23 mag.  The values of
$P(i)$ are a function of $I_{{\rm auto, lim}}$, decreasing to faint
magnitudes for early-type galaxies and increasing to faint magnitudes
for late types.  For those clusters with $I_{{\rm auto, lim}} > 22$
mag, we adopt the values of $P(i)$ down to $I_{\rm 814} = 22$ mag.
Based on the behavior of $P(i)$ versus $I_{\rm 814}$, this procedure
likely overestimates $P({\rm E})$, $P({\rm S0})$, and $P({\rm Sp})$
and underestimates $P({\rm Sp+Irr})$ and $P({\rm Irr})$.  These trends
translate into underestimates of the E, S0, and Sp morphological
fractions and overestimates of the Sp+Irr and Irr fractions.  These
mis-estimates will be most severe for cl1054-1245, cl1216-1201, and
cl1354-1230, which have the faintest values of $I_{{\rm auto, lim}}$.
The morphological fractions computed using statistical background
subtraction and Equation \ref{equation:statsub} are shown as hollow
triangles in Figure \ref{fig:MethodComparison}.  Examination of this
figure does not reveal any obvious biases with respect to other methods
that could be attributable to our $P(i)$ estimates.

As in \S \ref{sec:SpectroscopicRedshifts} and \S
\ref{sec:PhotometricRedshifts}, errors were computed using the
\citet{Gehrels86} approximation and N$_{\rm stat}(i)$ and N$_{\rm
stat}(tot)$.

\section{Results}
\label{sec:Results}

We have estimated the morphological fractions of 10 EDisCS clusters at
$0.5 < z < 0.8$ using four different background-subtraction techniques
(absolute limits from spectroscopic redshifts, direct estimates from
spectroscopic redshifts, photometric redshifts, and statistical
subtraction), within apertures of different radii (600 kpc and $0.6
\times R_{200}$), and for two different cosmologies (the classic
cosmology: $\Omega_0=1$, $\Lambda=0$, ${\rm H}_0 = 50$ km s$^{-1}$
Mpc$^{-1}$; and the WMAP cosmology: $\Omega_0=0.3$, $\Lambda=0.7$,
${\rm H}_0 = 70$ km s$^{-1}$ Mpc$^{-1}$ ).  Figure
\ref{fig:MethodComparison} shows how the different methods compare for
an aperture of 0.6 $\times R_{200}$ in the WMAP cosmology.  The
direct estimate methods are consistent with the absolute limits
derived from our spectroscopy, and with one another.  In addition,
there is no systematic trend for one estimate to produce higher or
lower fractions than any other.  For each cluster we therefore adopt a
single method: direct estimate from spectroscopy, photometric
redshifts, or statistical background subtraction.  The adopted method,
which differs from cluster to cluster, is chosen in the following way.
For a given cluster, median E, S0, and Sp+Irr fractions are selected
from among the estimates of all three methods.  The method selected
most often in this process is the one adopted for the cluster.  The E,
S0, and Sp+Irr fractions estimated using the chosen method for each
cluster are shown in Tables \ref{table:mergedtable1} and
\ref{table:mergedtable4}, and are the quantities plotted in Figures
\ref{fig:MorphologicalContent} through \ref{fig:sigmacorrelation}.

The morphological fractions in Figure \ref{fig:MethodComparison}
appear to be correlated with cluster number, which is determined by
right ascension.  Spearman and Kendall rank correlation tests show
that marginal correlations are detected at the $\sim$2$\sigma$ level
for the S0 and Sp+Irr fractions.  We find no correlation between
cluster redshift and cluster number that could explain this trend.
However, we find that the cluster velocity dispersion is correlated
with the cluster number, also at the $\sim$2$\sigma$ level.  Since
there is no reason for any physical property of the cluster to
correlate with cluster number, these correlations hint that the
morphological fractions correlate with cluster velocity dispersion, a
possibility we revisit in \S{\ref{sec:veldisp}}.

\subsection{Evolution of cluster morphological content}
\label{sec:Comparison}

To assess the degree of evolution in the morphological content of rich
clusters, we searched the literature for similar analyses of clusters
spanning a broad redshift range.  

The first comparison sample consists of $z<0.5$ clusters with
morphological fractions originally measured by different groups using
a variety of methods, but reanalyzed in a uniform manner by
\citet{Fasano00}, hereafter F00.  The F00 sample includes 55 low
redshift clusters \citep{Dressler80a, Dressler97}; the ten MORPHS
clusters at $0.37 < z < 0.5$ \citep{Dressler97}; three clusters at $z
\sim 0.3$ plus A2218 and A1689 at $z = 0.18$ \citep{Couch98}; and nine
clusters at $0.1 < z < 0.25$ \citep{Fasano00}.  Using this large
sample of uniformly analyzed clusters at $z<0.5$, \citet{Fasano00}
confirmed the MORPHs finding that the S0 fraction in clusters has
doubled since $z \sim 0.5$, while the late-type fraction has decreased
by a similar factor.

We also compare our results with that of \citet{Postman05}, who we
will refer to as P05.  They computed the morphological fractions
within the virial radii of 7 clusters at $0.8 < z < 1.27$.  This is
the only large study of clusters at these redshifts in which S0
galaxies are separately classified.  As previously discussed, the S0
population is an important one to track at high redshift.

In Figure \ref{fig:MorphologicalContent}, we show the evolution at $z
< 0.8$ of the morphological fractions within rich clusters, using
EDisCS data in conjunction with the F00 sample described above.  All
points were computed within 600 kpc, using the classic cosmology.  We
do not include measurements made on the Postman sample in this plot
because they were computed within $R_{200}$ using the WMAP cosmology.
As discussed previously, the F00 sample shows a systematic decrease in
the S0 fraction from 55\% at $z=0$ to 20\% at $z \sim 0.5$.  Where the
redshift range of the EDisCS clusters overlaps with that of F00, the
morphological fractions derived from the two samples are generally
consistent.  However, the longer redshift baseline afforded by the
addition of the EDisCS data reveals that the S0 fraction is actually
flat over the range $0.4 < z < 0.8$.  This could mean that $z \sim
0.4$ is a special epoch after which the S0 fraction in cluster cores
begins to grow.  The coincidence of this special epoch with the
redshift where the two samples overlap raises the question of whether
the samples were analyzed in different ways.  However, Section
\ref{sec:Analysis} describes our extensive attempts to control
systematics in the EDisCS sample and to conform to the analysis
presented by F00.  Alternatively, it is possible that we are missing
some S0-rich clusters at $z \sim 0.4$ which would reveal that the
growth of the S0 fraction in clusters at $z < 0.4$ is slower than
suggested by the present data.  Indeed, the scatter in the
morphological fractions at $z \sim 0.4$ is smaller than at either
higher or lower redshifts.

The F00+EDisCS data are consistent with either a bend in the
morphological fractions at $z \sim 0.4$ or a smooth continuation of
the trends observed at $z < 0.5$, but with a flatter overall slope
than suggested by the F00 data alone.  More observations at $z \sim
0.4$ and $z > 0.8$ are required to distinguish between these two
possibilities.  As discussed above, P05 have analyzed the
morphological fractions, including the S0 fraction, in clusters at $z
> 0.8$.  In Figure \ref{fig:MorphologicalContentpostman}, we show the
morphological fractions of the EDisCS and P05 clusters as functions of
redshift.  All estimates were made using the WMAP cosmology, but the
EDisCS fractions were computed within $0.6 \times R_{200}$ (see \S
\ref{sec:Aperture}), while the P05 fractions were computed within
$R_{200}$.  P05 find that the E and E+S0 fractions decrease out to
$0.6 \times R_{200}$, while the S0 fraction stays roughly constant and
the Sp+Irr fraction increases.  Between $0.6 \times R_{200}$ and
$R_{200}$, all the fractions are flat.  According to Table 4 of P05,
the E, S0, early, and late type fractions computed within $R_{200}$
are factors of approximately 0.79, 0.96, 0.85, and 1.15 times the
fractions computed within $0.6 \times R_{200}$.  However, no
corrections have been made to the points shown in Figure
\ref{fig:MorphologicalContentpostman}.

Figure \ref{fig:MorphologicalContentpostman} shows that the strong
evolution in morphological fractions seen at $z < 0.4$ does not
continue at $z > 0.4$.  Indeed, the E+S0 (Sp+Irr) fraction appears to
be larger (smaller) in the higher-redshift P05 sample compared to the
EDisCS clusters.  However, a Spearman rank correlation analysis shows
that there is no statistically significant evidence for any evolution
over the entire redshift interval $0.4 < z < 1.25$, or in the
individual EDisCS and P05 samples.  Additional clusters in this
redshift range are necessary to reveal the presence of any weak
correlation.

How does the selection requirement for the EDisCS clusters to exhibit
a red sequence affect the interpretation of Figures
\ref{fig:MorphologicalContent} and
\ref{fig:MorphologicalContentpostman}?  If the red sequence takes time
to build up, it may be expected that this requirement selects for
clusters that are also dynamically evolved.  In fact, the EDisCS
sample includes clusters that are clearly non-spherical, as well as
clusters that display significant substructure (see
\S{\ref{sec:veldisp}}).  The red sequence requirement may also be
expected to select for clusters with high E and/or S0 fractions.
Figure \ref{fig:MorphologicalContent} demonstrates that the EDisCS
clusters do not contain large E fractions compared to either the
optically-selected F00 sample at lower redshifts or the (primarily)
X-ray-selected P05 sample at higher redshifts.  Thus, either such a
bias is weak, or it is shared by the comparison samples.  Given the
correlation between galaxy color and morphology, any clusters lacking
a red sequence would likely have significantly different morphological
fractions from the EDisCS, F00, and P05 samples, leading to increased
scatter in Figures \ref{fig:MorphologicalContent} and
\ref{fig:MorphologicalContentpostman}.  While both the incidence of
such clusters and their morphological content are currently impossible
to quantify, in the next subsection we evaluate the dependence of
morphological fractions on another cluster property: velocity
dispersion.

\subsection{Correlation between Morphological Fractions and Cluster Velocity Dispersion}
\label{sec:veldisp}

In the previous subsection, we argued that there is no systematic
evolution in the morphological fractions within rich clusters at $0.4
< z < 1.25$.  Figures \ref{fig:MorphologicalContent} and
\ref{fig:MorphologicalContentpostman} indicate that the scatter in
these fractions is large.  How much of this scatter is due to a
correlation between morphological fractions and cluster mass?  Such a
correlation may be expected in either nature or nurture scenarios.  In
the former it is due to the fact that the most massive clusters
collapsed at earlier times.  In the latter it could be due to a higher
efficiency of morphological transformations in more massive clusters.

P05 found that the E, S0, and E+S0 fractions within $R_{200}$ of seven
$z \sim 1$ clusters increase with increasing bolometric cluster X-ray
luminosity, although the correlations are significant only at the
$\la$3$\sigma$ level.  However, they find no correlation between the
E+S0 fraction and the X-ray temperature or the cluster velocity
dispersion, perhaps because of small number statistics.  In Figure
\ref{fig:sigmacorrelation} we plot the morphological fractions of the
EDisCS and P05 samples against cluster velocity dispersion.  From this
figure it is apparent that clusters with larger velocity dispersions
harbor a higher fraction of early type galaxies and fewer late type
galaxies.  Spearman and Kendall rank correlation tests show that these
visual impressions are statistically significant at the $\la$3$\sigma$
level.  

While it is tempting to interpret this correlation as one between
morphological content and cluster mass, the velocity dispersion of a
cluster is directly related to its mass only if the cluster is
virialized.  The velocity dispersion may overpredict mass if the
cluster is experiencing significant merging events.  The degree of
substructure in five of the clusters in this work (cl1040-1155,
cl1054-1146, cl1054-1245, cl1216-1201, and cl1232-1250) has been
studied in detail by \citet{Halliday04} using Dressler-Shectman tests
\citep{Dressler88}.  They detect substructure in cl1232-1250 and
cl1216-1201 with more than 95\% confidence.  For cl1040-1155 and
cl1054-1245, not enough spectra were available to provide firm
evidence for substructure.  No evidence for substructure was found for
cl1054-1146.  They note that the two clusters showing clear evidence
of substructure also have the largest velocity dispersions in the
EDisCS sample, and caution against using the velocity dispersions for
these two clusters as a proxy for mass.  Further analysis on the
remaining EDisCS clusters is ongoing (Milvang-Jensen et al., in
preparation).

What is the situation for the $z< 0.5$ sample, where evolution is
observed?  Velocity dispersions were available in the literature for
14 of the clusters in the F00 sample
\citep{Couch87,Gudehus89,Girardi01,depropris02,Bettoni06}.  Figure
\ref{fig:sigmacorrelationlowz} shows the morphological fractions for
these 14 F00 clusters as a function of cluster velocity dispersion.
Without a larger number of data points it is difficult to make a
conclusive statement regarding the existence of a correlation.
Spearman and Kendall rank correlation tests indicate that none of the
morphological fractions are significantly correlated with cluster
velocity dispersion.  However, removal of two clusters (A389 and
A3330) with unusually high S0 fractions for their velocity dispersions
results in a 2$\sigma$ detection of a correlation between S0 fraction
and cluster velocity dispersion.  Additional velocity dispersions for
the F00 sample would greatly aid an assessment of any correlation,
which is necessary for understanding the observed trends between
morphology and redshift.

\begin{figure*}[h]
\epsscale{1.0}
\plotone{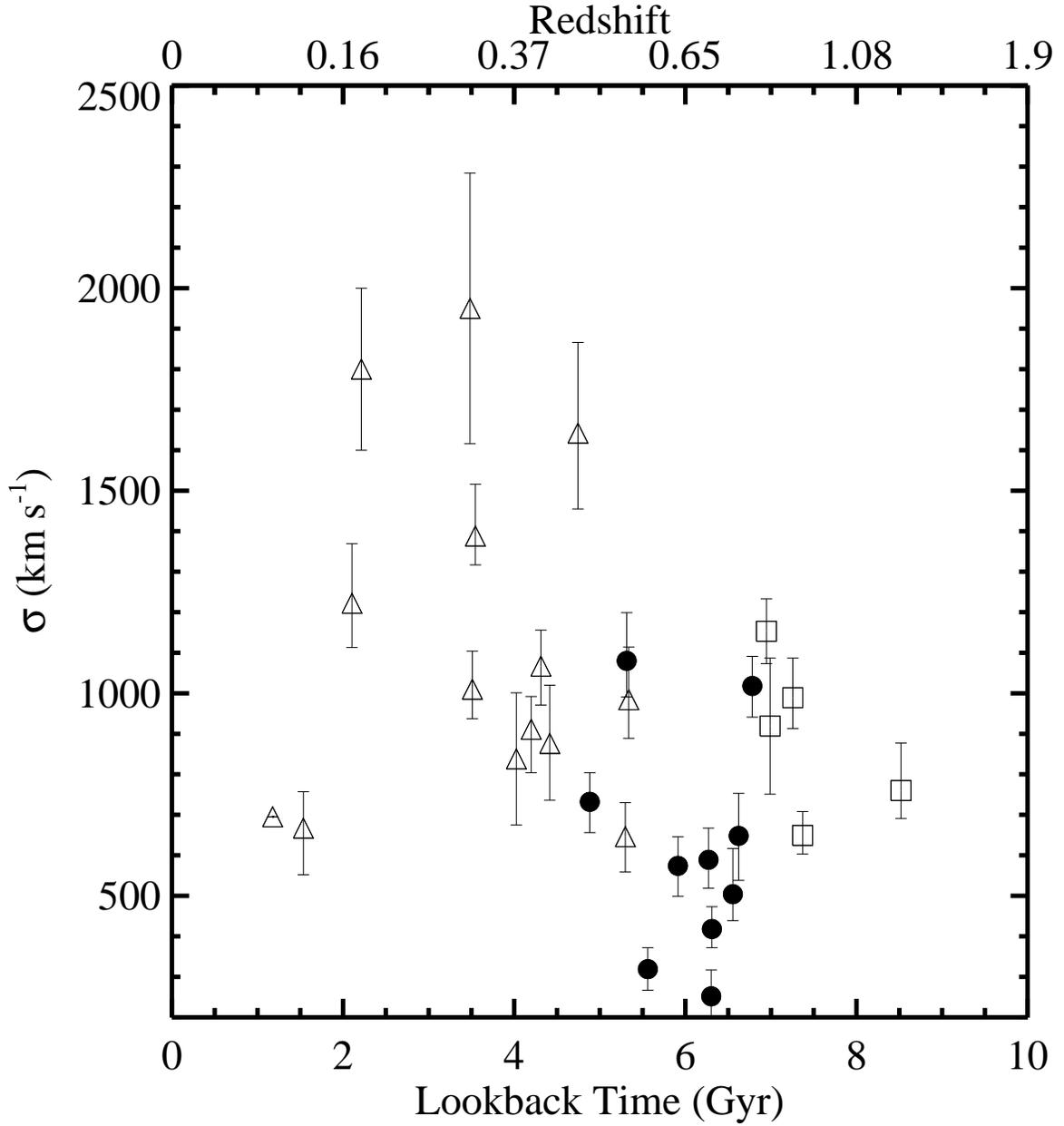}

\caption{Cluster velocity dispersion as a function of cluster lookback
time for the EDisCS sample (filled circles), the subset of the F00
sample for which velocity dispersions were available (hollow
triangles), and the subset of the P05 sample for which velocity
dispersions were available (hollow squares).  The lookback time was
calculated using the WMAP cosmology.}

\label{fig:sigma_v_redshift} 
\end{figure*}

Comparing the x-axes of Figures \ref{fig:sigmacorrelation} and
\ref{fig:sigmacorrelationlowz}, we see that the F00 sample at $z <
0.5$ includes some clusters with very high velocity dispersions
($\sigma > 1200$ km s$^{-1}$), while the $z > 0.5$ EDisCS+P05 sample
does not.  Figure \ref{fig:sigma_v_redshift} shows the cluster
velocity dispersions of the EDisCS, F00, and P05 samples as a function
of lookback time.  The clusters with the highest velocity dispersions
in these samples lie at low redshift.  Unfortunately, velocity
dispersions are unavailable for a significant fraction of the low
redshift sample.  These are urgently needed to understand the extent
to which sample selection is responsible for the apparent evolution in
the morphological content of clusters.

\section{Discussion and Conclusions}
\label{sec:Conclusions}

\begin{figure*}[h]
\epsscale{1.0}
\plotone{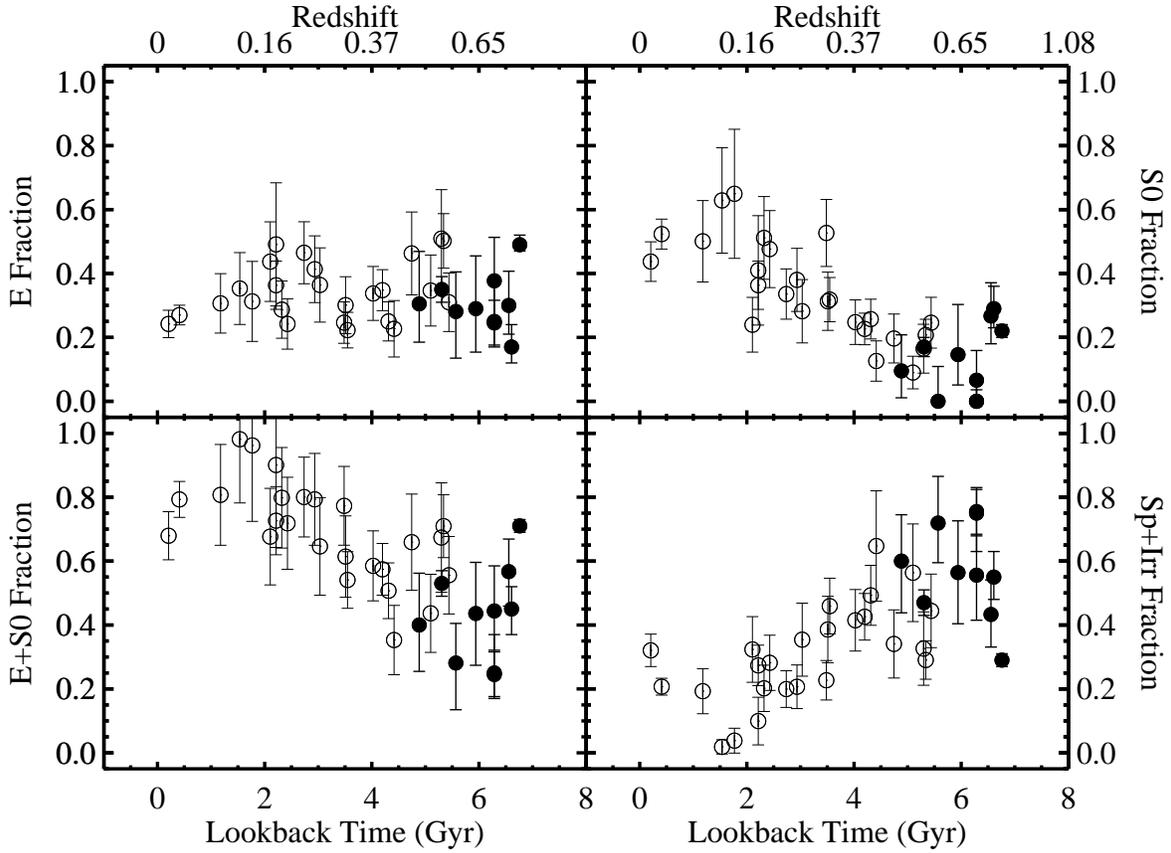}
\caption{The evolution of the E, S0, E+S0, and Sp+Irr fractions as
traced by EDisCS clusters (filled circles) and F00 clusters (hollow
circles). All fractions were computed within a radius of 600 kpc,
using the classic cosmology.  The lookback times were calculated with
the WMAP cosmology.}
\ref{fig:sigma_v_redshift}
\label{fig:MorphologicalContent} 
\end{figure*}

\begin{figure*}[h]
\epsscale{1.0}
\plotone{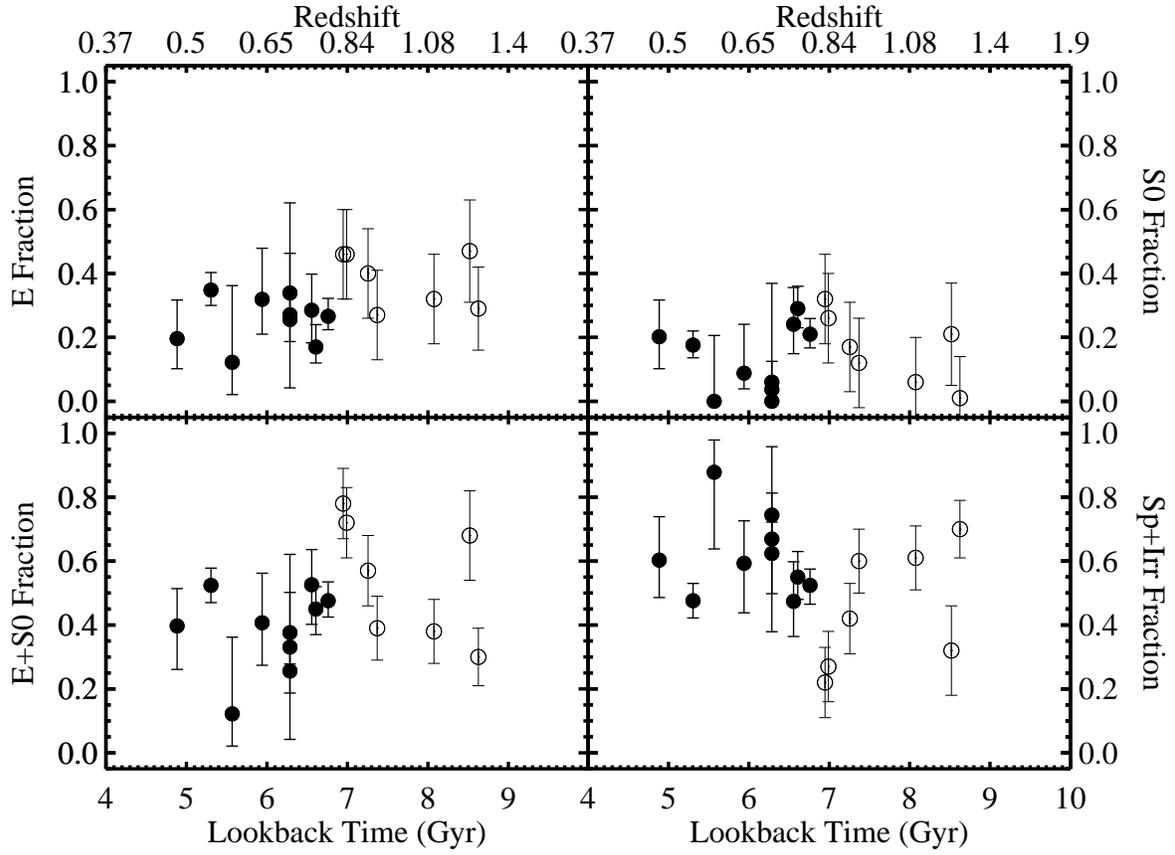}
\caption{The evolution of the E, S0, E+S0, and Sp+Irr fractions as
traced by EDisCS clusters (filled circles) and P05 clusters (hollow
circles).  The EDisCS fractions were computed within a radius of $0.6
\times R_{200}$ and the P05 fractions were computed within a radius of
$R_{200}$.  All computations were performed using the WMAP cosmology.}
\label{fig:MorphologicalContentpostman} 
\end{figure*}

Using high resolution imaging afforded by the ACS instrument on board
the Hubble Space Telescope, we have shown that the morphological
content of rich clusters with velocity dispersions in the range
$\sigma = 200-1200$ km s$^{-1}$ shows no evidence for evolution over
the redshift range $0.4 < z < 1.25$ (see Figures
\ref{fig:MorphologicalContent} and
\ref{fig:MorphologicalContentpostman}).  Although the scatter is
significant, typical morphological fractions for clusters in this
redshift range are 0.3, 0.15, and 0.55 for E, S0, and Sp+Irr galaxies,
respectively.  In contrast, studies of clusters at lower redshifts
have shown that the elliptical fraction remains constant between $z =
0$ and $z \sim 0.5$, while the S0 fraction decreases by a factor of
$\sim$2 and the Sp+Irr fraction increases by a similar amount over the
same redshift range (e.g. F00).  Our data show that the observed
evolution in cluster S0 and Sp+Irr populations does not continue
beyond $z \sim 0.4$, at least for the velocity dispersions probed in
this study.

How do our results concerning the global evolution of galaxy
morphologies within rich clusters translate into a statement regarding
the evolution of the morphology-density relation?  Although the
centers of cluster cores are regions of high local galaxy density, our
morphological fractions were computed within sizable apertures (see
\S \ref{sec:Aperture}), so the average environment we are probing is
of moderate density.  For example, the average galaxy surface density
within a radius of $0.6 \times R_{200}$ ($\Omega_0 = 0.3$, $\Lambda =
0.7$, H$_0 = 70$ km s$^{-1}$ Mpc$^{-1}$) for the EDisCS clusters used
in this analysis ranges from $\sim$40--175 Mpc$^{-2}$.  The
morphological fractions of the F00 and P05 clusters were likely
computed using galaxies inhabiting environments of similar average
density.  These results may point to a lack of evolution in the
morphology density relation between $z = 0.8$ and $z = 0.4$, with a
subsequent increase in the S0 population and a decrease in the Sp+Irr
population.  Larger samples in this redshift range are needed to
understand the scatter in the morphological fractions, and to rule out
weak evolution.  However, our current data are consistent with recent
studies of the morphology-density relation at $z \sim 1$
\citep{Smith05,Postman05}.

Several studies of nearby galaxies indicate that the observed relation
between morphology and environment is a reflection of a more primary
relation between star formation rate (SFR) and environment
\citep{Kauffmann04,Christlein05,Blanton05}.  The star formation
properties of cluster, group, and field galaxies in the EDisCS
spectroscopic sample have been measured and compared to a low redshift
sample from the Sloan Digital Sky Survey by \citet{Poggianti06} (P06).
Based on these data, P06 put forth a model in which passive galaxies
devoid of ongoing star formation are made up of two separate
populations: ``primordial'' galaxies whose stars all formed at $z >
2.5$ and galaxies whose star formation lasted until later times but
was ultimately ``quenched'' due to entering a cluster-like environment
($> 10^{14} {\rm M}_{\odot}$).  In this model, the primordial galaxies
may be identified with the elliptical galaxies that make up $\sim$30\%
on average of the galaxy populations in clusters at $z < 1$, and
perhaps some of the S0 galaxies (see below).  This identification is
consistent with both the lack of evolution in the elliptical fraction
at $z < 1$ and with the old ages inferred for elliptical stellar
populations.  It is tempting to identify the quenched galaxies with
S0s.  However, the fraction of quenched galaxies in the model
increases at $z<0.8$, while the S0 fraction only increases at $z <
0.4$.  These timescales are consistent if it takes roughly a billion
years for a galaxy to resemble an S0 after the cessation of star
formation in a spiral galaxy \citep{Dressler99, Poggianti99, Tran03}.
However, the formation of the S0 population observed in clusters at
$0.4 < z < 1$ cannot be explained in this way.  Perhaps these are
``primordial'' S0s that assumed an S0 morphology upon or soon after
formation at $z > 2.5$.  If so, the S0 galaxies in clusters at $z >
0.4$ should be, on average, older and more massive than cluster S0s at
$z < 0.4$.  Another possibility is that at least some late types may
transform into S0s in systems less massive than $10^{14} M_{\odot}$,
perhaps by an entirely different mechanism than operates in $>10^{14}
M_{\odot}$ systems.  The existence of S0s in groups
\citep[e.g.][]{Hickson89} suggests that this may be the case.

If a significant fraction of S0 galaxies in nearby clusters were
quenched, what did they look like beforehand?  The parallel decline in
the Sp+Irr fraction as the S0 fraction increases between $z=0.4$ and
$z=0$ suggests that some subset of late type galaxies transformed into
S0s subsequent to quenching.  For the EDisCS clusters, we computed the
fractions of Sp and Irr galaxies separately (see Tables
\ref{table:mergedtable1} through \ref{table:mergedtable4}).  As with
the other morphological fractions at $z > 0.4$, the Sp and Irr
fractions do not vary systematically with redshift.  Moreover, the Irr
fractions in the EDisCS clusters are very small.  We conclude that Sp
galaxies and not Irr galaxies are the precursors of the quenched S0s
in local clusters.

In the P06 model, both nature and nurture (within clusters) play a
role in the environmental dependence of star formation and therefore
morphology.  The nurture component within clusters is consistent with
the observation that the relation between star formation and
environment in nearby clusters is not solely a reflection of the
relation between stellar age and environment, as would be expected in
a pure nature scenario \citep{Christlein05}.  The P06 model is also
compatible with observations which show that the SFR-density relation
out to $z \sim 1$ extends to very low local densities, comparable to
those found at the virial radius of clusters and even outside clusters
\citep{Lewis02, Gomez03,Cooper06}.  The low density relations could be
set up by the dependence of galaxy properties on initial conditions,
or perhaps a different mechanism creates S0 galaxies in low-density
environments, or a combination.

Although the model forwarded by P06 is an attractive framework within
which to view our observational results on the morphological content
of rich clusters, it is not necessarily the only model which could
explain all of the data accumulating on the evolution of galaxy
properties with environment.  Furthermore, it does not identify a
mechanism responsible for the ``quenching'', although it does specify
a timescale (3 Gyr) and a mass scale ($10^{14} M_{\odot}$) of a
workable mechanism.  
  
\begin{figure*}[h]
\epsscale{1.0}
\plotone{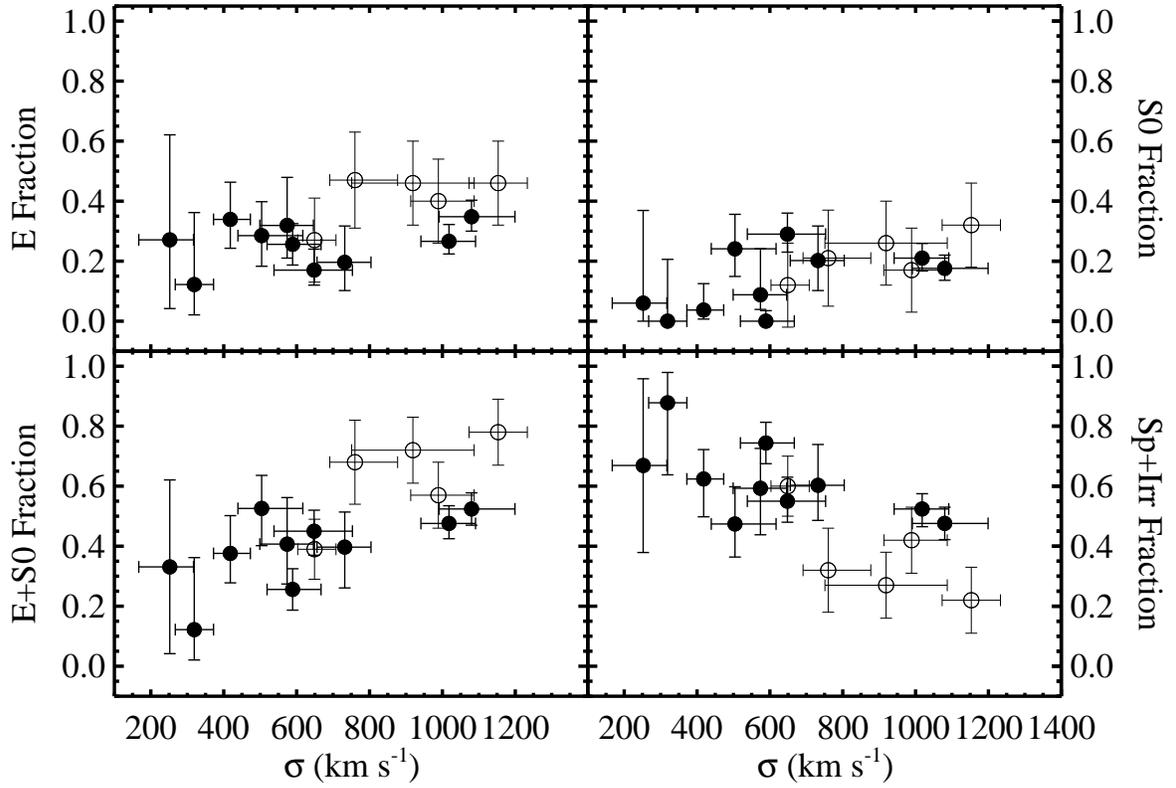}
\caption{Morphological fractions as a function of cluster velocity
dispersion for the EDisCS clusters (filled circles) and the subset of
the P05 cluster sample for which velocity dispersions are available
(hollow circles).  The EDisCS fractions were computed within a radius
of $0.6 \times R_{200}$ and the P05 fractions were computed within a
radius of $R_{200}$.  All fractions were computed using the WMAP
cosmology.}
\label{fig:sigmacorrelation} 
\end{figure*}

\begin{figure*}[h]
\epsscale{1.0}
\plotone{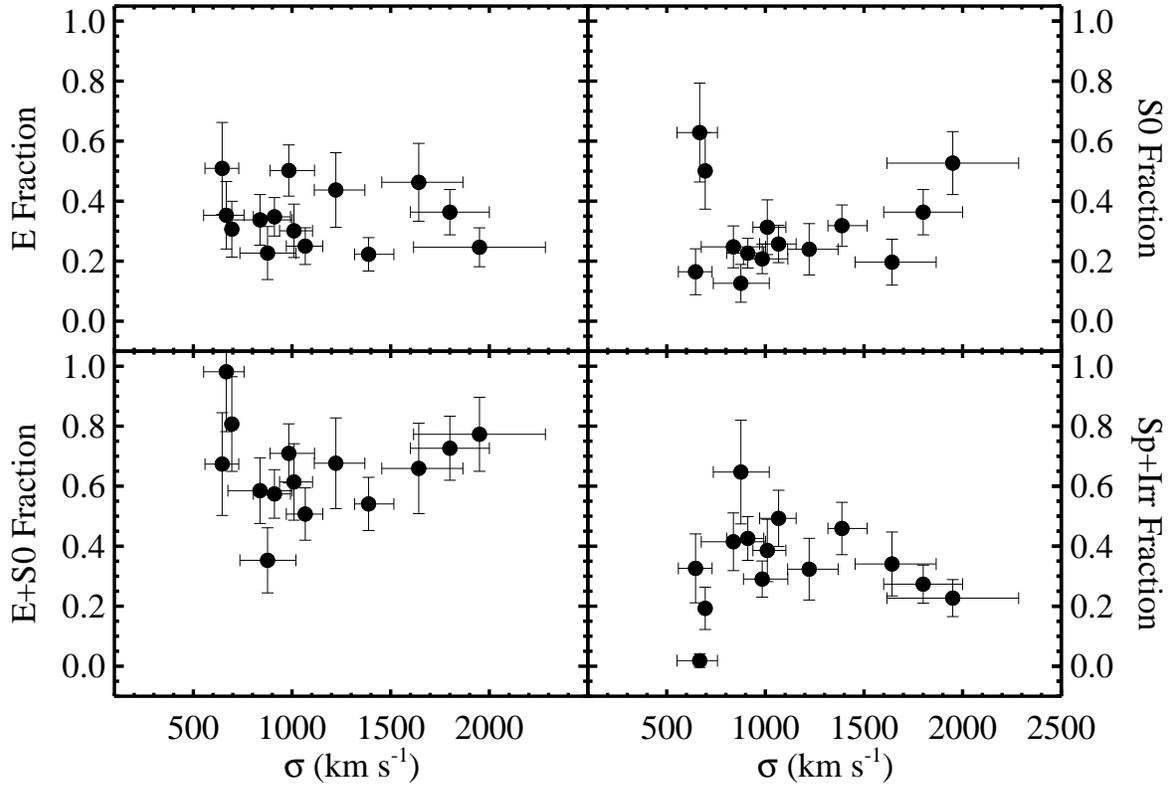}

\caption{Morphological fractions as a function of cluster velocity
dispersion for the subset of the F00 cluster sample for which velocity
dispersions are available.  All morphological fractions were computed
within 600 kpc using the classic cosmology.  The velocity dispersion
measurements for ten of these clusters come from \citet{Girardi01}.
The remainder come from \citet{Bettoni06} (A3330 with $\sigma = 695$
km s$^{-1}$), \citet{depropris02} (A389 with $\sigma = 667$ km
s$^{-1}$), \citet{Gudehus89} (A1689 with $\sigma = 1800$ km s$^{-1}$),
and \citet{Couch87} (AC118 with $\sigma = 1950$ km s$^{-1}$).  Both
A3330 and A389 have large S0 fractions compared to other clusters with
similar velocity dispersions.  In addition, A1689 and AC118 are the
F00 clusters with the largest velocity dispersions, and largely drive
the correlation between morphological fractions and velocity
dispersion.  This plot illustrates the need for homogeneous studies of
velocity dispersions for clusters where evolution in morphological
fractions is observed.}

\label{fig:sigmacorrelationlowz} 
\end{figure*}

In addition to analyzing the evolution of the morphological content in
galaxy clusters out to $z = 1.25$, we have also studied the dependence
of the morphological content on the cluster velocity dispersion (see
Figures \ref{fig:sigmacorrelation} and
\ref{fig:sigmacorrelationlowz}).  We find that the early and late-type
fractions in the EDisCS and P05 cluster samples correlate with cluster
velocity dispersion at a statistically significant level.  This
correlation highlights the importance of understanding global cluster
properties in samples where evolution is observed, such as the F00
sample.  Unfortunately, only limited velocity dispersion information
is available for the F00 clusters.  The existing information for 14
clusters indicates that they have a higher average velocity dispersion
than either the EDisCS or P05 clusters.  In addition, the
morphological fractions of these 14 clusters do not correlate strongly
with cluster velocity dispersion.  It is possible that additional
velocity dispersions for the F00 sample will reveal an underlying
trend, that additional studies at high redshift will prove the trend
observed in the higher redshift EDisCS+P05 sample to be spurious, or
that the relation between morphological fraction and cluster velocity
dispersion is itself a function of redshift.  Larger cluster samples
with robust velocity dispersions and morphologies are needed to
determine which of these options is most likely.  If the
morphology-density relation is driven by environmental processes in
clusters, such measurements are essential for determining how the
efficiency of these transformations depends on cluster velocity
dispersion, which, as discussed in \S{\ref{sec:veldisp}} is related to
both the cluster mass and its dynamical state.

\acknowledgments

We greatly appreciate the timely and cheerful assistance of Galina
Soutchkova in planning the HST observations.  We also thank the
referee, Manolis Plionis, for feedback which improved the paper.  VD
acknowledges funding from the Graduate Student Researchers Program.
This work was supported by NASA grant HST-GO-09476.01.  JJD was
partially supported by the Alfred P.\ Sloan Foundation.  The Medium
Deep Survey catalog is based on observations with the NASA/ESA Hubble
Space Telescope, obtained at the Space Telescope Science Institute,
which is operated by the Association of Universities for Research in
Astronomy, Inc., under NASA contract NAS5-26555. The Medium-Deep
Survey analysis was funded by the HST WFPC2 Team and STScI grants
GO2684, GO6951, GO7536, and GO8384 to Prof. Richard Griffiths and Dr
Kavan Ratnatunga at Carnegie Mellon University.

\bibliographystyle{apj}
\bibliography{references.bib}

\clearpage

\begin{deluxetable}{ccccccccc}[h]
  \tablecaption{The EDisCS HST Sample}
  \tabletypesize{\footnotesize}
  \tablewidth{0pt}
  \tablecolumns{8}
  \tablehead{
    \colhead{Number} &
    \colhead{Name} & 
    \colhead{RA (J2000)} & 
    \colhead{Dec (J2000)} & 
    \colhead{$z$} & 
    \colhead{$\sigma$} & 
    \colhead{$R_{200,1}$} & 
    \colhead{$R_{200,2}$} \\ 
    \colhead{} & 
    \colhead{} & 
    \colhead{(h:m:s)} & 
    \colhead{(\degr:\arcmin:\arcsec)} & 
    \colhead{} & 
    \colhead{(km s$^{-1}$)} & 
    \colhead{(Mpc)} & 
    \colhead{(Mpc)} & 
    \colhead{}}
  \startdata
1 &   cl1037-1243  & 10:37:51.4 & -12:43:26.6          & 0.58 & $319_{-52}^{+53}$   &     0.56  &  0.57 \\
2 &   cl1040-1155  & 10:40:40.3 & -11:56:04.2          & 0.70 & $418_{-46}^{+55}$   &     0.65  &  0.70 \\
3 &   cl1054-1146  & 10:54:24.4 & -11:46:19.4          & 0.70 & $589_{-70}^{+78}$   &     0.92  &  0.99 \\
4 &   cl1054-1245  & 10:54:43.5 & -12:45:51.9          & 0.75 & $504_{-65}^{+113}$  &     0.75  &  0.82 \\
5 &   cl1103-1245b & 11:03:36.5 & -12:44:22.3          & 0.70 & $252_{-85}^{+65}$   &     0.39  &  0.42 \\
6 &   cl1138-1133  & 11:38:10.2 & -11:33:37.9          & 0.48 & $732_{-76}^{+72}$   &     1.41  &  1.40 \\
7 &   cl1216-1201  & 12:16:45.3 & -12:01:17.6          & 0.79 & $1018_{-77}^{+73}$  &     1.47  &  1.61 \\
8 &   cl1227-1138  & 12:27:58.9 & -11:35:13.5          & 0.64 & $574_{-75}^{+72}$   &     0.95  &  1.00 \\
9 &   cl1232-1250  & 12:32:30.3 & -12:50:36.4          & 0.54 & $1080_{-89}^{+119}$ &     1.95  &  1.99 \\
10 &  cl1354-1230  & 13:54:09.8 & -12:31:01.5          & 0.76 & $648_{-110}^{+105}$ &     0.96  &  1.05 \\
  \enddata
  \tablecomments{The RA, Dec coordinates are the positions of the
    brightest cluster galaxy.  The redshifts ($z$) and line-of-sight
    cluster velocity dispersions ($\sigma$) were taken from
    \citet{Halliday04} and Milvang-Jensen et al.~(in preparation).  The
    values of the virial radius, $R_{200}$, listed in columns 6 and 7
    were computed using Equation 1 in \citet{Poggianti06} using
    the classic cosmology and the WMAP cosmology, respectively, as described in
    \S\ref{sec:intro}.}
  \label{table:HSTSample}
\end{deluxetable}

\begin{deluxetable}{lp{12.0cm}}
  \tablecaption{Notes on Parameters in Morphological Catalogs.}
  \tablewidth{0pt}
  \tablehead{\colhead{Heading}   & \colhead{Description}}
  \startdata
  ID                     & EDisCS ID\\
  RA                     & Right Ascension in decimal degrees (J2000)\\
  DEC                    & Declination in decimal degrees (J2000)\\
  $I_{\rm auto}$         & Total $I$ magnitude; SExtractor MAG\_AUTO parameter measured from ground-based $I$-band images\\
  Type                   & star=-7, non-stellar but too compact to see structure=-6, E=-5, S0=-2, Sa=1, Sb=3, Sc=5, Sd=7, Sm=9, Irr=11, no HST data corresponding to ground-based object=111, unclassifiable=66\\
  S0 disk flag           & If any of the classifiers noted that the B/D ratio is small but the disk is featureless, this flag is 1.\\
  bar flag               & If any of the classifiers noted the presence of a bar, this flag is 1.\\
  edge-on flag           & If any of the classifiers noted that this galaxy is edge-on, this flag is 1.\\
  small flag             & If any of the classifiers noted that this galaxy is small, this flag is 1.\\
  LSB flag               & If any of the classifiers noted that this galaxy is low-surface-brightness, this flag is 1.\\
  defect flag            & If any of the classifiers noted that the image of this galaxy was defective, due for example to cosmic rays or incomplete coverage, this flag is 1.\\
  dust flag              & If any of the classifiers noted the presence of dust in the galaxy, this flag is 1.\\
  disturbance flag       & If any of the classifiers noted that this galaxy is disturbed, this flag is 1.\\
  comments               & Additional comments by any of the classifiers\\
  \enddata
  \label{table:parameters}
\end{deluxetable}

\begin{deluxetable}{ccccccccccc}
  \tablecaption{Morphological Fractions within EDisCS Clusters:  $R=600$ kpc, $\Omega_0 = 1$, $\Lambda = 0$, H$_0 = 50$ km s$^{-1}$ Mpc$^{-1}$}
  \tablewidth{0pt}
  \tabletypesize{\footnotesize}
  \tablehead{
    \colhead{Cluster Name}   & 
    \colhead{$f_{\rm E}$} & 
    \colhead{$f_{\rm S0}$} & 
    \colhead{$f_{\rm E+S0}$} & 
    \colhead{$f_{\rm Sp+Irr}$} & 
    \colhead{$f_{\rm Sp}$} & 
    \colhead{$f_{\rm Irr}$} }
  \startdata
cl1037-1243  &  0.281$^{+0.124}_{-0.146}$  &  0.000$^{+0.109}_{-0.000}$  &  0.281$^{+0.124}_{-0.146}$  &  0.719$^{+0.146}_{-0.124}$  &  0.625$^{+0.138}_{-0.156}$  &  0.095$^{+0.174}_{-0.050}$ \\
cl1040-1155  &  0.377$^{+0.136}_{-0.116}$  &  0.066$^{+0.093}_{-0.058}$  &  0.444$^{+0.141}_{-0.123}$  &  0.556$^{+0.123}_{-0.141}$  &  0.419$^{+0.141}_{-0.115}$  &  0.137$^{+0.127}_{-0.071}$ \\
cl1054-1146  &  0.245$^{+0.071}_{-0.069}$  &  0.000$^{+0.036}_{-0.000}$  &  0.245$^{+0.071}_{-0.069}$  &  0.755$^{+0.069}_{-0.071}$  &  0.755$^{+0.069}_{-0.071}$  &  0.000$^{+0.036}_{-0.000}$ \\
cl1054-1245  &  0.300$^{+0.107}_{-0.090}$  &  0.267$^{+0.104}_{-0.087}$  &  0.567$^{+0.102}_{-0.108}$  &  0.433$^{+0.108}_{-0.102}$  &  0.433$^{+0.108}_{-0.102}$  &  0.000$^{+0.060}_{-0.000}$ \\
cl1103-1245b  &  0.250$^{+0.120}_{-0.080}$  &  0.000$^{+0.070}_{-0.000}$  &  0.250$^{+0.120}_{-0.080}$  &  0.750$^{+0.080}_{-0.120}$  &  0.750$^{+0.080}_{-0.120}$  &  0.000$^{+0.070}_{-0.000}$ \\
cl1138-1133  &  0.305$^{+0.164}_{-0.120}$  &  0.095$^{+0.113}_{-0.084}$  &  0.400$^{+0.162}_{-0.145}$  &  0.600$^{+0.145}_{-0.162}$  &  0.600$^{+0.145}_{-0.162}$  &  0.000$^{+0.115}_{-0.000}$ \\
cl1216-1201  &  0.490$^{+0.030}_{-0.020}$  &  0.220$^{+0.020}_{-0.020}$  &  0.710$^{+0.020}_{-0.020}$  &  0.290$^{+0.020}_{-0.020}$  &  0.270$^{+0.020}_{-0.020}$  &  0.020$^{+0.010}_{-0.010}$ \\
cl1227-1138  &  0.290$^{+0.165}_{-0.136}$  &  0.146$^{+0.157}_{-0.095}$  &  0.436$^{+0.160}_{-0.162}$  &  0.564$^{+0.162}_{-0.160}$  &  0.394$^{+0.167}_{-0.164}$  &  0.170$^{+0.132}_{-0.120}$ \\
cl1232-1250  &  0.350$^{+0.040}_{-0.040}$  &  0.170$^{+0.030}_{-0.030}$  &  0.530$^{+0.040}_{-0.040}$  &  0.470$^{+0.040}_{-0.040}$  &  0.470$^{+0.040}_{-0.040}$  &  0.000$^{+0.010}_{-0.000}$ \\
cl1354-1230  &  0.170$^{+0.070}_{-0.050}$  &  0.290$^{+0.070}_{-0.060}$  &  0.450$^{+0.070}_{-0.080}$  &  0.550$^{+0.080}_{-0.070}$  &  0.550$^{+0.080}_{-0.070}$  &  0.000$^{+0.030}_{-0.000}$ \\
  \enddata
  \label{table:mergedtable1}

  \tablecomments{The morphological fractions for a given cluster were
  computed using the background subtraction method adopted for that
  cluster (see \S{\ref{sec:Results}}).}

\end{deluxetable}

\begin{deluxetable}{ccccccccccc}
  \tablecaption{Morphological Fractions within EDisCS Clusters:  $R=0.6 \times R_{200}$, $\Omega_0 = 0.3$, $\Lambda = 0.7$, H$_0 = 70$ km s$^{-1}$ Mpc$^{-1}$}
  \tablewidth{0pt}
  \tabletypesize{\footnotesize}
  \tablehead{
    \colhead{Cluster Name}   & 
    \colhead{$f_{\rm E}$} & 
    \colhead{$f_{\rm S0}$} & 
    \colhead{$f_{\rm E+S0}$} & 
    \colhead{$f_{\rm Sp+Irr}$} & 
    \colhead{$f_{\rm Sp}$} & 
    \colhead{$f_{\rm Irr}$} }
  \startdata
cl1037-1243  &  0.122$^{+0.240}_{-0.101}$  &  0.000$^{+0.206}_{-0.000}$  &  0.122$^{+0.240}_{-0.101}$  &  0.878$^{+0.101}_{-0.240}$  &  0.854$^{+0.124}_{-0.217}$  &  0.023$^{+0.182}_{-0.023}$ \\
cl1040-1155  &  0.339$^{+0.124}_{-0.096}$  &  0.037$^{+0.088}_{-0.030}$  &  0.376$^{+0.126}_{-0.098}$  &  0.624$^{+0.098}_{-0.126}$  &  0.624$^{+0.098}_{-0.126}$  &  0.000$^{+0.068}_{-0.000}$ \\
cl1054-1146  &  0.256$^{+0.069}_{-0.069}$  &  0.000$^{+0.035}_{-0.000}$  &  0.256$^{+0.069}_{-0.069}$  &  0.744$^{+0.069}_{-0.069}$  &  0.744$^{+0.069}_{-0.069}$  &  0.000$^{+0.035}_{-0.000}$ \\
cl1054-1245  &  0.285$^{+0.113}_{-0.102}$  &  0.241$^{+0.115}_{-0.092}$  &  0.526$^{+0.110}_{-0.124}$  &  0.474$^{+0.124}_{-0.110}$  &  0.469$^{+0.108}_{-0.120}$  &  0.005$^{+0.066}_{-0.005}$ \\
cl1103-1245b  &  0.271$^{+0.350}_{-0.229}$  &  0.060$^{+0.309}_{-0.060}$  &  0.331$^{+0.290}_{-0.289}$  &  0.669$^{+0.289}_{-0.290}$  &  0.669$^{+0.289}_{-0.290}$  &  0.000$^{+0.369}_{-0.000}$ \\
cl1138-1133  &  0.196$^{+0.121}_{-0.094}$  &  0.202$^{+0.115}_{-0.100}$  &  0.397$^{+0.117}_{-0.136}$  &  0.603$^{+0.136}_{-0.117}$  &  0.603$^{+0.136}_{-0.117}$  &  0.000$^{+0.084}_{-0.000}$ \\
cl1216-1201  &  0.266$^{+0.056}_{-0.042}$  &  0.210$^{+0.049}_{-0.043}$  &  0.476$^{+0.059}_{-0.051}$  &  0.524$^{+0.051}_{-0.059}$  &  0.432$^{+0.053}_{-0.055}$  &  0.092$^{+0.036}_{-0.031}$ \\
cl1227-1138  &  0.319$^{+0.160}_{-0.109}$  &  0.088$^{+0.153}_{-0.049}$  &  0.407$^{+0.155}_{-0.133}$  &  0.593$^{+0.133}_{-0.155}$  &  0.494$^{+0.146}_{-0.135}$  &  0.099$^{+0.142}_{-0.059}$ \\
cl1232-1250  &  0.348$^{+0.055}_{-0.048}$  &  0.176$^{+0.044}_{-0.040}$  &  0.524$^{+0.054}_{-0.054}$  &  0.476$^{+0.054}_{-0.054}$  &  0.425$^{+0.052}_{-0.054}$  &  0.051$^{+0.029}_{-0.024}$ \\
cl1354-1230  &  0.170$^{+0.070}_{-0.050}$  &  0.290$^{+0.070}_{-0.060}$  &  0.450$^{+0.070}_{-0.080}$  &  0.550$^{+0.080}_{-0.070}$  &  0.550$^{+0.080}_{-0.070}$  &  0.000$^{+0.030}_{-0.000}$ \\
  \enddata
  \label{table:mergedtable4}
  \tablecomments{The morphological fractions for a given cluster were
  computed using the background subtraction method adopted for that
  cluster (see \S{\ref{sec:Results}}).}
\end{deluxetable}

\end{document}